\documentclass[aps,pra,twocolumn,floatfix,showpacs,noshowkeys,epsfig,graphics]{revtex4}
\usepackage{amsmath,amssymb,bm,graphicx}

\newcommand{\ket}[1]{|{#1} \rangle}

\begin{document}
\title{Entanglement and Quantum Phase Transitions via Adiabatic Quantum Computation}
\author{Sangchul Oh}~\email{oh.sangchul@gmail.com}
\affiliation{Department of Physics, University at Buffalo, The State University of New York, 
       Buffalo, New York 14260-1500, USA}

\begin{abstract}
For a finite XY chain and a finite two-dimensional Ising lattice, it is shown that 
the paramagnetic ground state is adiabatically transformed to the GHZ state 
in the ferromagnetic phase by slowly turning on the magnetic field. The fidelity 
between the GHZ state and an adiabatically evolved state shows a feature of 
the quantum phase transition.
\end{abstract}
\pacs{03.65.Ud, 03.67.Lx, 03.67.Mn, 73.43.Nq, 75.10.Jm}
%Entanglement \sep XY model \sep GHZ state \sep Quantum phase transition
%03.65.Ud: Entanglement and Non-locality
%03.67.Lx: Quantum Computation
%03.67.Mn: Entanglement production, characterization
%05.70.Fh: phase transition, General study
%73.43.Nq: Quantum phase transitions, 
%75.10.Jm: Quantized Spin model
\maketitle

\section{Introduction}
A quantum computer is a quantum system, so it could simulate quantum dynamics
directly and efficiently~\cite{Feynman82}. When a quantum system undergoes 
a quantum phase transition (QPT) \cite{Sachdev99}, induced by 
the variation of external parameters or coupling strength, its ground state 
changes dramatically and maybe, too, its entanglement. Entangled states, 
showing quantum correlations between subsystems, are not only valuable 
resources in quantum information processing but also important for 
understanding quantum many-body systems. So much attention has been paid 
to a study of entanglement of quantum many-body systems in ground states or 
at thermal equilibrium in connection with 
QPTs~\cite{Osterloh02,Osborne02,Amico08}. However, the simulation of 
QPTs and the generation of entangled states 
with quantum computers are less explored.

In this paper, we address whether QPTs can be simulated with adiabatic 
quantum computation (AQC) and how the entanglement changes when 
the QPT takes place. We present a way to generate an entangled state 
of a quantum system which undergoes a QPT during the quantum adiabatic 
evolution. As prototypes of QPTs, we consider a spin $1/2$ XY chain and 
a two-dimensional spin $1/2$ Ising lattice. It is shown that 
a product state in the paramagnetic phase is adiabatically transformed to 
a Greenberg-Horne-Zeilinger (GHZ) entangled state, in the ferromagnetic 
phase, and {\it vice versa}. We shows the fidelity between the GHZ state 
and an adiabatically evolved state could be a good indicator to QPTs. 
For a two-dimensional Ising model, {\it a two-dimensional} GHZ state is
generated via AQC.

%\section{Entanglement and QPTs via Adiabatic Quantum Computation}
\section{Adiabatic Quantum Computation}
Let us start with a brief introduction to AQC~\cite{Farhi01}. Quantum
computation can be implemented by the controlled dynamics of quantum states 
governed by the Schr\"odinger equation with a time-dependent Hamiltonian 
\begin{equation}
i\hbar\frac{d}{dt}\ket{\Psi(t)} = H(t) \ket{\Psi(t)} \,,
\label{Schroedinger}
\end{equation}
and quantum measurements. In a quantum circuit model, 
the evolution of a quantum state is decomposed into a series of single-qubit 
and two-qubit gates, which can be implemented by applying 
external pulses and by controlling the interaction between two qubits.
On the other hand, AQC relies on the adiabatic theorem, which states that an 
evolved quantum system will stay at its instantaneous eigenstate if the 
time-dependent Hamiltonian changes very slowly. If the Hamiltonian $H(t)$ in 
Eq.~(\ref{Schroedinger}) changes slowly, then the initial state $\ket{\Psi(0)}$,
taken as an eigenstate state $\ket{\varphi_n(0)}$ of an initial Hamiltonian $H(0) 
\equiv H_0$ at $t=0$, evolves to $\ket{\Psi(T)} =\ket{\varphi_n(T)}$, an 
eigenstate of a problem Hamiltonian $H(T) \equiv H_P$ at $t=T$. 
The run time $T$ is inversely proportional to the square of the minimum energy gap 
during the evolution. It is convenient to introduce the dimensionless time 
$s \equiv t/T$ with $0\le s \le 1$. There are many ways to connect $H_0$ and $H_P$ 
smoothly as a function of $s$, for example, simple linear or nonlinear 
interpolations~\cite{Farhi02}. A general interpolation is given by $H(s) = f(s) H_0 
+ g(s) H_P$ where two functions $f(s)$ and $g(s)$ satisfy the
boundary conditions, $f(0) = g(1) = 1$ and $f(1) = g(0) =0$.
It is known that a proper interpolation could reduce the run time of AQC. 

\section{Entanglement and QPTs of the XY chain via AQC}
The Hamiltonian of a spin $1/2$ XY chain in a transverse 
magnetic field is written as
\begin{align}
H_{XY} =& -\sum_{i=1}^{N} \left[ 
          \left( \frac{1+\gamma}{2} \right) \sigma_{i}^z\sigma_{i+1}^z
        + \left( \frac{1-\gamma}{2} \right) \sigma_{i}^y\sigma_{i+1}^y
          \right] \nonumber\\
        &-\lambda\sum_{i=1}^{N} \sigma_{i}^{x} \,,
\label{Hamil_xy}
\end{align}
where $N$ is the total number of spins, $\lambda$ the transverse magnetic field, 
and $\gamma$ the parameter for the degree of anisotropy of spin-spin 
interaction. Here, the coordinates $x$ and $z$ are exchanged for convenience as in
Ref.~\cite{Sachdev99}. The periodic boundary condition, $\sigma_{N+1} =
\sigma_{1}$, is assumed. For $\gamma =1$, it becomes the Ising model 
$H_I = -\sum_{i=1}^{N}\sigma_{i}^z\sigma_{i+1}^z -\lambda\sum_i \sigma_{i}^{x}$.
For $\gamma = 0$, it is called the XX model.

Let us recall the ground state of the Ising model which undergoes 
the QPT at $\lambda_c =1$~\cite{Sachdev99}. If $\lambda\gg 1$, 
the Zeeman term in Eq.~(\ref{Hamil_xy}) is dominant and
the ground state is given by the product of eigenstates of $\sigma_{i}^{x}$,
called the paramagnetic state
\begin{equation}
\ket{P} \equiv \prod_{i=1}^{N} \ket{+}_i\,, 
\label{para_state}
\end{equation}
where $\ket{+} = \frac{1}{\sqrt{2}} 
\left( \ket{\uparrow} + \ket{\downarrow} \right)$. 
In the other limit of $\lambda\ll 1$, the spin-spin interaction in Eq.~(\ref{Hamil_xy})
is important and the ground state has two-fold degeneracy. 
A possible ground 
state can be any superposition of all spin up state $\ket{F_\uparrow}$ and all 
spin down state $\ket{F_\downarrow}$ where 
\begin{equation}
\ket{F_\uparrow} \equiv \prod_{i=1}^{N} \ket{\uparrow}_i\,, \quad
\ket{F_\downarrow} \equiv \prod_{i=1}^N \ket{\downarrow}_i \,.
\label{ferro_state}
\end{equation}
One possible ground state is the GHZ state of $N$ spins 
\begin{equation}
\ket{\rm GHZ}_N = \frac{1}{\sqrt{2}}(\ket{F_\uparrow} + \ket{F_\downarrow}) \,.
\label{GHZ_state}
\end{equation}
In the thermodynamic limit, $N\to \infty$, the ground could be
either $\ket{F_\uparrow}$ or $\ket{F_\uparrow}$ due to the spontaneous 
symmetry breaking. However, a computer resource is {\it finite} no matter whether
it is classical or quantum. So we focus on the adiabatic quantum simulation 
of QPTs and the generation of entangled states with a finite system.

The spin $1/2$ XY chain is exactly solvable in the sense that its energy spectrum, 
ground state, and phase diagram can obtained via the mapping of spin operators to
fermion operators via the Jordan-Wigner transformation~\cite{Lieb61,Barouch71}. 
Recently, this system has attracted much attention in the study of the relation 
between entanglement and QPTs. Osborne and Nielsen~\cite{Osborne02} studied the 
entropy of a single spin and the two-spin entanglement, where $\ket{F_{\uparrow}}$ 
was  taken as a ground state in the ferromagnetic phase. 
Osterloh {\it et al.}~\cite{Osterloh02} investigated the two spin entanglement 
of the spin $1/2$ XY chain and showed that the concurrence as a two-spin 
entanglement measure exhibits 
the characteristic features of QPTs. Note that both $\ket{F_{\uparrow}}$ 
and $\ket{\rm GHZ}_N$ have the same value of the two-spin entanglement, i.e., zero 
concurrence, although $\ket{P}$, $\ket{F_\uparrow}$, and $\ket{F_\downarrow}$ are 
separable states but $\ket{\rm GHZ}_N$ is entangled. The behavior of multi-particle 
entanglement at QPTs is an open problem because a good entanglement measure for 
more than two spins is still under development~~\cite{Wu04,Yang05,Venuti06,Oliveira06}. 
The direct simulation of QPTs via AQC might give a clue to this problem. 

\begin{figure}[htbp]
\centering{\includegraphics[scale=0.8]{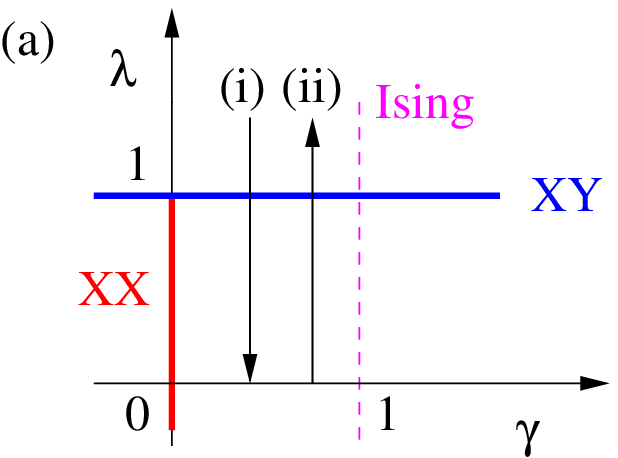}} \\
\centering{\includegraphics[scale=1.0]{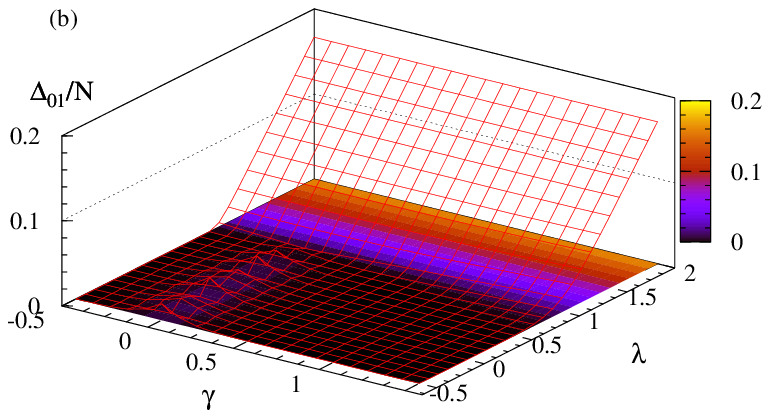}}
\caption{(a) Phase diagram of the XY chain with the XY critical line 
       at $\lambda = 1$ and the XX critical line at $\gamma =0$~\cite{Barouch71}. 
       It is symmetric about $\gamma$ and $\lambda$ axes.
      Path (i) starts from the paramagnetic phase and end at the ferromagnetic phase. 
      The return path is denoted by (ii). (b) Energy gap $\Delta_{01}/N$ between the two lowest 
      eigenvalues as a function of $\gamma$ and $\lambda$.}
\label{Fig1}
\end{figure}

For the simulation of QPTs of the XY model via AQC, let us decompose the Hamiltonian $H_{XY}$ 
into two parts: the initial Hamiltonian $H_0$ and the problem Hamiltonian  $H_P$
\begin{equation}
H_0 = -\sum_{i=1}^N \sigma_{i}^{x} \,,\quad
H_P(\gamma) = H_{XY} -\lambda H_0\,.
\end{equation}
Among various ways of connecting $H_0$ and $H_P$, two interpolation schemes, 
linear and square ones, are considered  to see whether a proper 
interpolation could reduce the run time of AQC. The linear interpolation 
is given by
\begin{eqnarray}
H_{XY}(s,\gamma) &= (1-s)H_0 +  sH_P(\gamma)\,, \label{Hamil_lin}
\end{eqnarray}
and the square interpolation reads 
\begin{eqnarray}
H_{XY}(s,\gamma) &= (1-s^2)H_0 + s(2-s) H_P(\gamma)\,, \label{Hamil_sq}
\end{eqnarray}
By comparing of Eqs.~(\ref{Hamil_xy}), (\ref{Hamil_lin}), and (\ref{Hamil_sq}), 
one obtains the time-dependence of the magnetic field $\lambda(s) = (1-s)/s$ for 
the linear interpolation, and $\lambda(s) = (1-s^2)/(2s-s^2)$ for 
the square interpolation. The path of adiabatic evolution from 
the paramagnetic phase to the ferromagnetic phase is depicted in 
Fig.~\ref{Fig1} (a). At time $s=0$ corresponding to the limit 
$\lambda \to\infty$, the initial state is given by the paramagnetic state,
Eq.~(\ref{para_state}). At $s=1/2$, i.e., $\lambda =1$, the system arrives at 
the XY critical line. The adiabatic evolution ends at $s=1$, that is, $\lambda = 0$.
We examine which of two states, Eqs.~(\ref{ferro_state}) and~(\ref{GHZ_state}) 
is the true final state by calculating the fidelity between the GHZ state and
an evolved state as function of $s$ and $\gamma$. Note that Wei 
{\it et al.}~\cite{Wei05} used the maximum fidelity between a state and 
an untangled state as a global entanglement measure in the study of 
the multi-particle entanglement of a XY chain.
To check the reversibility of AQC, we investigate the reverse path from 
the ferromagnetic phase, starting with 
(\ref{ferro_state}) or~(\ref{GHZ_state}), to the paramagnetic phase by 
exchanging $H_0$ and $H_P$.

For the numerical simulation of the AQC, we develop the program which solves 
the Schr\"odinger equation and diagonalizes the Hamiltonian directly 
without the Jordan-Wigner transformation. We simulate the spin $1/2$ XY 
chain with $N=12$, and a two-dimensional Ising model of size $3\times 3$ 
on a personal computer.

For the one-dimensional XY model, the energy spectrum is obtained 
as function of $\lambda$ and the anisotropy parameter $\gamma$ as shown in 
Fig.~\ref{Fig1} (b). As the free energy determines classical phase 
transitions, the energy gap between the ground state and the first 
exited state plays a key role in QPTs.  
As depicted in Fig.~\ref{Fig1} (b), the gap $\Delta_{01}$ between 
the two lowest eigenvalues of the XY Hamiltonian clearly vanishes at
the critical line, i.e., at $\lambda =1$. On this line the two lowest eigenvalues 
merge together and the ground state becomes degenerate, even though 
the system size is finite. 

\begin{figure}[htbp]
\centering{\includegraphics[scale=1.0]{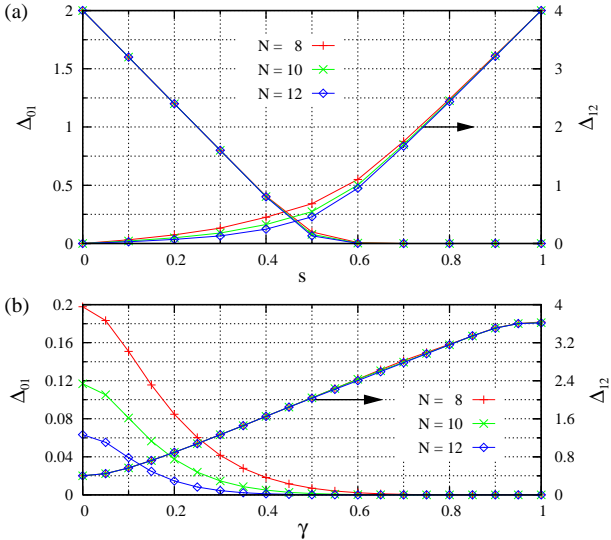}}
\caption{(a) $\Delta_{01}$ and $\Delta_{12}$ as a function of $s$ 
  at $\gamma =1$ for $N=8,10,12$. (b) $\Delta_{01}$ and $\Delta_{12}$
  as a function of $\gamma$ at $\lambda =0.1$ for $N=8, 10,12$. 
  The linear interpolation is used.}
\label{Fig2}
\end{figure}

Fig.~\ref{Fig2} (a) plots
$\Delta_{01}$ and $\Delta_{12}$, the gap between the second and third 
lowest eigenvalues, as a function of $s$ at $\gamma =1$ for $N=8,10,12$.
$\Delta_{01}$ is nearly independent of $N$. However, $\Delta_{12}$ is more
dependent on $N$ near the critical points $s_c=1/2$, i.e., $\lambda_c = 1$. 
The energy gap $\Delta$ of the spin $1/2$ XY chain with infinite size
is known to be $\Delta \sim |\lambda - \lambda_c|$~\cite{Sachdev99}.
$\Delta_{01}$ and $\Delta_{12}$ can be regarded as the energy gaps between 
the ground state and the first excited state in the paramagnetic phase 
and in the ferromagnetic phase, respectively. 
That is $\Delta(\lambda) = \Delta_{01}(\lambda)$ for $\lambda > 1$ and 
$\Delta(\lambda) = \Delta_{12}(\lambda)$ for $\lambda <1$.
So the critical slowing down at the critical point~\cite{Zurek05} is due 
to $\Delta_{12}$ not due to $\Delta_{01}$ as $N\to \infty$. 
Let us examine the universality of $\Delta_{01}$, that is, 
independent of $\gamma$. Due to the finite size effect, the region of the 
universality defined by $\Delta_{01} =0$ does not extend to the XX line, 
i.e., $0 <\gamma \le 1$ for the infinite lattices.
The region satisfying the universality grows with $N$ as shown in 
Fig.~\ref{Fig2} (b).

\begin{figure}[htbp]
\centering{\includegraphics[scale=1.0]{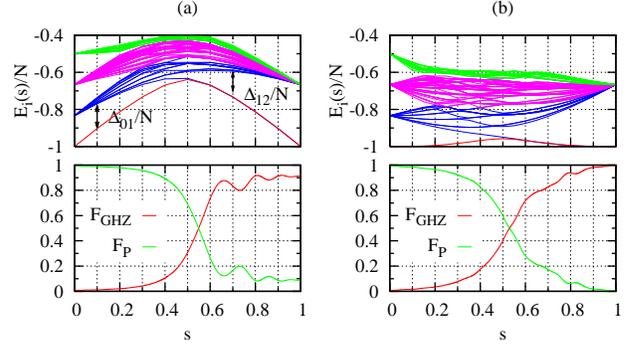}}
\caption{ Energy levels per spin and the fidelity as 
   a functions of $s$ for the XY model with $N=12$ and $\gamma =0.75$ 
   (a) for the linear interpolation, and (b) for the square interpolation. 
   The run time $T=20$ is taken. The energy spectrum ranges from -1 to 1 
   and is symmetric about $x$ axis.}
\label{Fig3}
\end{figure}

Fig.~\ref{Fig3} shows the difference between the linear interpolation and 
the square interpolation. The linear interpolation needs more run time $T$, 
than the square one. The main reason is that the gap $\Delta_{12}$ for the square 
interpolation is larger at the critical region than that for the linear one, 
so the probability for the transition to the excited states is reduced. 
We find that the square interpolation could reduce the run time $T$ of AQC in 
the case of the XY chain. 

\begin{figure}[htbp]
\centering{\includegraphics[scale=1.0]{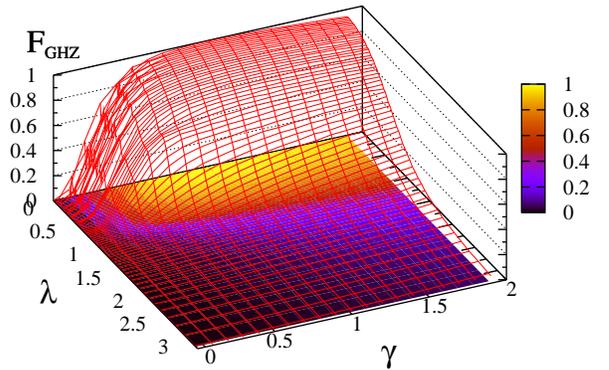}}
\caption{Fidelity $F_{\rm GHZ}$ between the GHZ state and 
         the evolved state as a function of $\lambda(s)$ and $\gamma$. 
         $N=10$, $T=20$, and the square interpolation 
         are used.}
\label{Fig4}
\end{figure}

Fig.~\ref{Fig4} shows the fidelity between an evolved state and the GHZ state, 
$F_{\rm GHZ}(s) =|\langle\Psi(s)|{\rm GHZ}\rangle|^2$, as a function of $s$ 
and $\gamma$.  The paramagnetic state $\ket{P}$ in Eq.~(\ref{para_state})
at $s=0$ ($\lambda \gg 1$) adiabatically evolves to the GHZ state
$\ket{\rm GHZ}_N$ in Eq.~(\ref{GHZ_state}) 
at $s=1$ ($\lambda=0$). Especially, at $\gamma = 1$, this result is consistent 
with Dorner {\it et al.}'s one ~\cite{Dorner03}. They showed that for the Ising chain 
the paramagnetic state is transformed to the GHZ state by slowly decreasing the magnetic 
field $\lambda$. Although due to the finite size effect, the fidelity decreases 
near the XX critical line, it is almost independent of of $\gamma$. This is one of 
the characteristic features of QPTs, called the universality.
As mentioned before, the region of the universality is dependent on the number 
of spins $N$. 

\begin{figure}[htbp]
\centering{\includegraphics[scale=1.0]{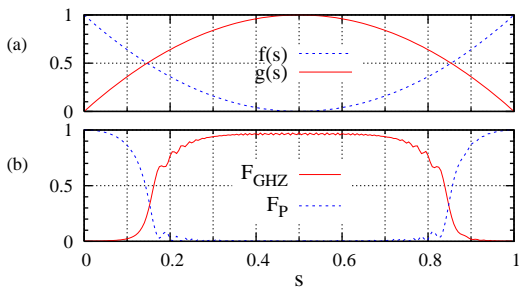}}
\caption{(a) Two functions $f(s)$ and $g(s)$
    as a function $s$ for the round trip.
   (c) $F_{\rm GHZ}$ and $F_P$ as a function of $s$ with $\gamma=0.8$, $N=10$, $T=200$.}
\label{Fig5}
\end{figure}

Let us discuss the reversibility of the QPT. The paramagnetic state is 
adiabatically transformed to the GHZ state in the ferromagnetic phase 
by decreasing $\lambda$. Does the GHZ state evolve adiabatically 
to the paramagnetic state (\ref{para_state}) even though there is 
the energy level splitting at the XY critical line?
By exchanging $H_0$ and $H_P$, the reverse evolution, the return path (ii) in 
Fig.~\ref{Fig1} (a), can be implemented. To examine the reversibility 
of AQC, let us consider $H_{XY}(s,\gamma) = f(s)H_0 + g(s)H_P(\gamma)$ where 
$f(s)=4(s-{1}/{2})^2$ and $g(s)=-4s(s-1)$. At $s=(2-\sqrt{2})/4$, 
the paramagnetic to ferromagnetic transition happens.  The ferromagnetic 
to paramagnetic transition takes place at $s=(2+\sqrt{2})/4$.
We find that in spite of the energy level merging and splitting during 
the journey, the paramagnetic state is adiabatically transformed to the GHZ 
state and {\it vice versa} as shown in Fig.~\ref{Fig5} (b).

Let us discuss the effect of finite system size on the AQC simulation of QPTs and 
the generation of entangled states. Although the system considered here is finite, 
it could show important features of QPTs. First, QPTs are characterized by the energy 
level crossing or avoided crossing~\cite{Sachdev99}. Fig.~\ref{Fig1} (b) shows 
energy level crossings at $\lambda =1$. Second, a $XY$ spin chain exhibits 
the universality, which state that the physics is independent of $\gamma$ 
for $0<\gamma\le 1$.
The energy gap in Figs.~\ref{Fig1} (b) and the fidelity in Fig.~\ref{Fig4} are
independent of $\gamma$, although the universality is not perfect due to the 
finite size effects. Third, Fig.~\ref{Fig4} shows the abrupt change in the 
ground state at QPT. One unsolved issue is that the entanglement of the ground
state in the ferromagnetic phase. If either $\ket{F_\uparrow}$ or 
$\ket{F_\downarrow}$ is the ground state in the ferromagnetic phase,
there is no entanglement. On the other hand, the ground state $\ket{\rm GHZ}_N$ is 
a multi-particle entangled state. In this case,
the single-site entropy varies from $0$ at the paramagnetic phase to $1$ in 
the ferromagnetic phase. This is in contrast with the result of~\cite{Osborne02}. 
A related question is whether the entanglement of the ground state in ferromagnetic 
or paramagnetic phase is constant or dependent on $\lambda$. In other 
words, does entanglement change abruptly only at the critical point? 

\section{Entanglement of two-dimensional Ising model}

\begin{figure}[ht]
\centering{\includegraphics[scale=0.5]{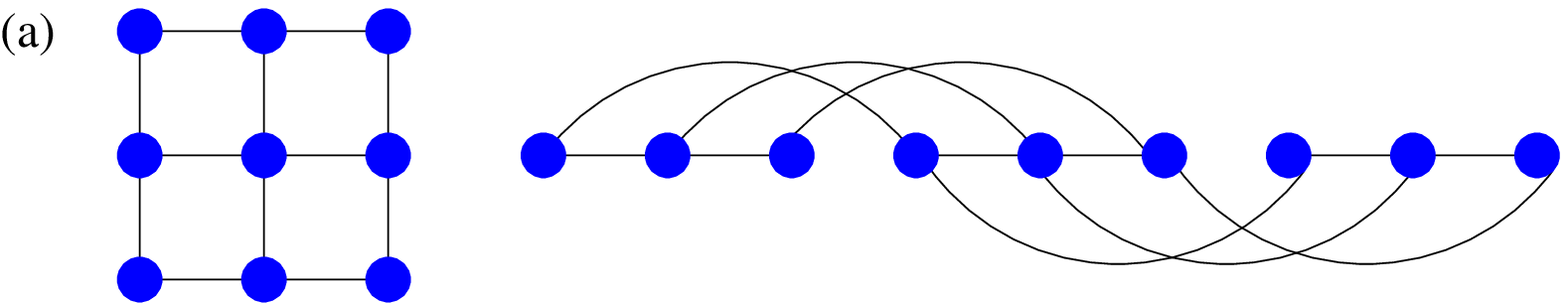}}\\
\centering{\includegraphics[scale=1.0]{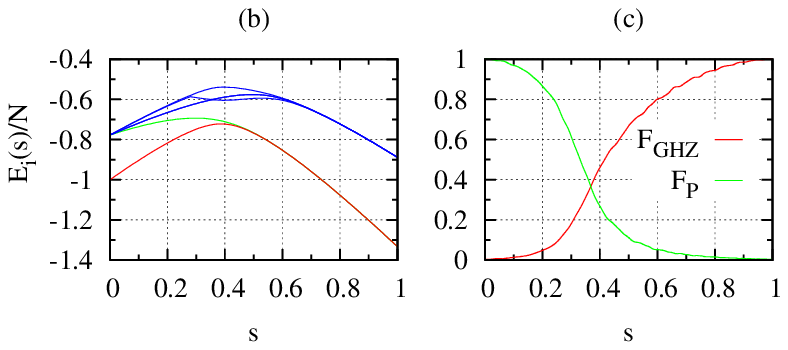}}
\caption{Two dimensional square lattice of size $3\times 3$ 
   and its mapping to the one-dimensional lattice with long-range interaction.
   (b) Energy levels as a function of $s$ with the linear interpolation. 
   (c) Fidelity between the evolved state and the two-dimensional GHZ state 
   as a function of $s$.}
\label{Fig:2D_Ising}
\end{figure}

Let us turn to the two-dimensional Ising model 
to produce the two-dimensional GHZ state 
\begin{equation}
\ket{\rm GHZ}_{2D} =\frac{1}{\sqrt{2}}\left( \prod_{i,j=1}^{N} \ket{\uparrow}_{ij}
           +\prod_{i,j=1}^{N} \ket{\uparrow}_{ij} \right)\,,
\label{2D_GHZ}
\end{equation}
where $\ket{\uparrow}_{ij}$ is the spin-up state at the lattice site $i$ and $j$.
As illustrated in Fig.~\ref{Fig:2D_Ising} (a), the $3\times 3$ two-dimensional lattice 
is considered. 
The open boundary condition is assumed. The two-dimensional Ising model can be 
mapped to the one-dimensional Ising model with long-range interactions as 
depicted in Fig.~\ref{Fig:2D_Ising} (a). Fig.~\ref{Fig:2D_Ising} (b) and (c) show
several lowest energy levels, and the fidelity between an evolved state and 
the two-dimensional GHZ state defined by Eq. (\ref{2D_GHZ}) as a function of $s$.
Like one-dimensional spin 1/2 XY model, the two dimensional GHZ state can be 
generated via AQC.

\section{Summary}
We have considered the one-dimensional XY model and the two-dimensional Ising 
model and simulated the QPTs and the generation of an entangled state 
via AQC. Although the system size is finite, our results show the characteristic 
features of QPTs. It has been demonstrated that the paramagnetic state evolves 
adiabatically to the GHZ state in the ferromagnetic phase and vice versa. 
The generation of entangled states via AQC is simple in the sense 
that only the external magnetic field is turned off or on slowly.  
It doesn't require the control of the exact qubit-qubit coupling, i.e., 
CNOT gate. We have shown that a square interpolation scheme is better in reducing
the run time than linear one. 

One open issue in AQC is that the minimum energy gap, which determine the 
run time $T$, should be known before running. Also the run time should 
be smaller than the decoherence time but at the same time large enough to 
avoid the unwanted transition. We are studying the effect of decoherence on 
AQC by solving the Lindblad master equation~\cite{Oh02} and the generation of 
W-type entangled states or cluster states~\cite{Briegel01} via AQC. 
Also, it is interesting to study how to simulate a quantum system in the 
thermodynamic limit with finite quantum computational resources. 

\section*{Acknowledgments}
The main part of this work was done when S. Oh was at Korea Institute for 
Advanced Study, Seoul, Korea.

\end{document}